\documentclass[preprints,article,accept,pdftex,moreauthors]{Definitions/mdpi} 
\firstpage{1} 
\pubvolume{1}
\issuenum{1}
\articlenumber{0}
\pubyear{2023}
\copyrightyear{2023}
\externaleditor{Academic Editor: Vladimir I. Manko}
\datereceived{15 December 2022} 
\daterevised{18 January 2023} 
\dateaccepted{25 January 2023} 
\datepublished{}
\hreflink{https://doi.org/} 

\graphicspath{{./figures/}}

\usepackage{braket,physics}
\usepackage{dsfont}
\usepackage[labelformat=simple]{subcaption}

\DeclareCaptionLabelFormat{subcaptionlabel}{\normalfont(\textbf{#2}\normalfont)}
\usepackage[inline]{trackchanges}

\newcommand{\inlineText}[2]{\footnotesize\UserLabel~#2}
\let\oldmarginText\marginText

\makeatletter
\renewcommand{\marginText}[2]{%
  \@ifundefined{@captype}{\oldmarginText{#1}{#2}}{\inlineText{#1}{#2}}%
}
\makeatother

\Title{Coarse-Grained Effective Hamiltonian via the Magnus Expansion for a Three-Level System}

\TitleCitation{Coarse-Grained Effective Hamiltonian via the Magnus Expansion for a Three-Level System}

\Author{Nicola Macrì
 $^{1,2,}$*\orcidA{}, Luigi Giannelli $^{1,3}$\orcidB{}, Elisabetta Paladino $^{1,2,3}$\orcidD{} and Giuseppe Falci $^{1,2,3}$\orcidC{}}

\AuthorNames{Nicola Macrì, Luigi Giannelli, Elisabetta Paladino and Giuseppe Falci}

\AuthorCitation{Macrì, N.; Giannelli, L.; Paladino, E.; Falci, G.}

\address{%
$^{1}$ \quad Dipartimento di Fisica e Astronomia “Ettore Majorana”, Università di Catania, Catania, 95123, Italy; luigi.giannelli@dfa.unict.it (L.G.); elisabetta.paladino@dfa.unict.it (E.P.); giuseppe.falci@unict.it (G.F.)
\\
$^{2}$ \quad INFN, Sezione di Catania, 95123 Catania, Italy\\
$^{3}$ \quad CNR-IMM, UoS Università, 95123 Catania, Italy\\}

\corres{Correspondence: nicola.macri@dfa.unict.it}

\abstract{Quantum state processing is one of the main tools of quantum technologies. While real systems are complicated and/or may be driven by non-ideal control, they may nevertheless exhibit simple dynamics  approximately confined to a low-energy Hilbert subspace. Adiabatic elimination is the simplest approximation scheme allowing us to derive in certain cases an effective Hamiltonian operating in a low-dimensional Hilbert subspace.  
However, these approximations may present ambiguities and difficulties, hindering a systematic improvement of their accuracy in larger and larger systems. Here, we use the Magnus expansion as a systematic tool to derive ambiguity-free effective Hamiltonians. We show that the validity of the approximations ultimately leverages only on a proper  coarse-graining in time of the exact dynamics. We validate the accuracy of the obtained effective Hamiltonians with suitably tailored fidelities of quantum operations.
}


\begin{document}



\section{Introduction}
In the last decade, research in quantum physics is experiencing a second
quantum revolution~\cite{dowling2003quantum, deutsch2020harnessing}. Huge
efforts have been and are being made in developing and engineering quantum
hardware and control~\cite{KochEQT2022quantum}, allowing novel quantum tasks to be performed in computation~\cite{nielsen2002quantum},
communication~\cite{gisin2007quantum,ka:09-benentifalci-prl-memorych}, and
sensing~\cite{degen2017quantum,bongs2019taking}. When dealing with real-life
quantum hardware, one possibly has to take into account the presence of states
that are not populated during the dynamics but still affect it via virtual
processes. {They form a subspace whose dimensions may become exponentially large with the size of the system, and they may produce various phenomena from the renormalization of coupling constants to leakage}~\cite{ka:16-theis-pra-simgatesfreqcrowded} {from the relevant ``computational'' Hilbert subspace}.
These non-relevant sectors can be removed {or ``sterilized''} if an effective Hamiltonian $\hat H_\mathrm{eff}$ is determined that describes the same dynamics of the original $\hat H$ for relevant cases, and it is of course much simpler than the original one. Effective models are key tools in the study of complex quantum systems~\cite{ka:22-li-prx-nonperturbativeandiag} since a simpler Hamiltonian~\cite{GoerzNJP2014optimal} may allow for analytical solutions or for faster convergence of numerical calculations~\cite{BLANES2009151}. 

Adiabatic elimination (AE) is perhaps the simplest and still successful {state-based} method to determine $\hat H_\mathrm{eff}$, eliminating states that are mostly not populated from the dynamics. However, this approach presents ambiguities and limitations (see Section~\ref{sec:AE-ambiguities}) pointed out, for instance, in Refs.~\cite{ka:07-brion-jphysA-adiabelimination,ka:14-paulish-epjplus-adiabelim}, where they have been tackled using various techniques such as Green's functional formalism~\cite{ka:07-brion-jphysA-adiabelimination} and exploiting Markov approximation in a Lippman--Schwinger approach~\cite{ka:14-paulish-epjplus-adiabelim}. 
A canonical {Hamiltonian-based} method for computing low-energy effective Hamiltonians relies on the Schrieffer--Wolff transformation technique~\cite{bravyi2011schrieffer,bukov2016schrieffer}, i.e., searching for a suitable unitary transformation that approximately decouples  the relevant and the non-relevant subspaces.

Here, we propose a derivation of an effective coarse-grained Hamiltonian, which is naturally free from the ambiguities and limitations mentioned above.  
The method is based on the Magnus expansion (ME), which expresses the solution of a differential equation in an exponential form~\cite{blanes2010pedagogical,BLANES2009151}. Applied to the time-evolution operator $U$ in a suitable coarse-graining time $\tau$, it yields an approximation whose logarithm gives an effective Hamiltonian. In general, the result depends on $\tau$ which must be chosen, if possible, in a proper way depending on the values of the system parameters and the significance of the relevant dynamics obtained a posteriori. As a benchmark, we apply the method to a three-level system in Lambda configuration. We discuss the systematic improvement of the approximation and compare it with other strategies. We check the validity of the results by using various figures of merit, the long-time fidelity of quantum evolution being the most informative one. 

\section{Methods}
\subsection{Effective Hamiltonian by the Magnus Expansion}
\label{sec:effH-magnus}
The Magnus expansion is a mathematical tool that allows one to express the solution of a differential equation in an exponential form. We apply that to the time-evolution operator $ U(t, t_0) $ of a quantum system that solves the Schr\"odinger equation $ \dot{U}(t,t_0) = - i \, H(t) \, U(t,t_0) $ for initial condition $ U(t_0,t_0) = \mathds{1} $. The Magnus expansion allows us to write the logarithm of $U(t,t_0)$ as a series, or, likewise,

$$
U(t,t_0) = \mathrm{e}^{ -i \, \sum_i F_i(t,t_0) }
$$

The two lowest-order terms of the expansion are given by {($\hbar =1$)}

\begin{equation}
\label{eq:magnus-terms}
\begin{aligned}
	F_1 &= \int_{t_0}^t \, ds \, H(s),
	\\
	F_2 &= -\frac{i}{2} \int_{t_0}^t \int_{t_0}^{s_1} ds_1 \, ds_2 \, \comm{H(s_1)}{H(s_2)}
\end{aligned}
\end{equation}

with $ s_1 > s_2 $. The term $F_1$ yields the so-called average Hamiltonian~\cite{PetiziolSR2020optimized}, while $F_2$
provides already an excellent approximation in most cases. Higher-order terms
involve time-ordered integrals of higher-order nested
commutators~\cite{blanes2010pedagogical,BLANES2009151,vandersypen2005nmr} and
are reported in Appendix~\ref{app:ME_high_order}.

Coarse-graining of the dynamics is operated by first splitting the evolution operator into time slices $ \tau = (t-t_0)/N $

$$
U(t, t_0) = \prod_j U \Big(t_j \, + \,\frac{\tau}{2} , t_j \, - \frac{\tau}{2} \Big) =: \prod_j U_j.
$$

Then we consider the ME in the $ j $-th sub-interval and truncate the series at a given order $n$

{ \protect $$
	i\,\ln U_j \approx
	i \, \sum_{i=1}^n F_i(t_j \, + \,\frac{\tau}{2} , t_j \, - \frac{\tau}{2}) =: H_\mathrm{eff}(t_j|\tau) \,\tau
	$$}
	
\noindent \textls[25]{obtaining an effective Hamiltonian (for a brief discussion about convergence, see} \mbox{Appendix~\ref{app:convergence}}). The structure of the ME suggests that accuracy is related to the smallness of the commutator $[H(t), H(t^\prime)]$ at different times, and it may be increased by choosing a small enough $\tau$.{ At the same time, if a $\tau$ that is large enough can be chosen, then the fast dynamics in the integrals defining $F_i$ is averaged out.} When successful, this procedure defines a coarse-grained Hamiltonian $H_\mathrm{eff}(t_j)$ whose explicit dependence on $\tau$ {can be neglected}.{ We} finally obtain the approximate time-evolution operator as 

{\protect $$
	U(t,t_0) \approx \prod_j \mathrm{e}^{- i H_\mathrm{eff}(t_j) \tau} \approx {\mathcal T}
	\mathrm{exp} \Big\{- i \int_{t_0}^t ds \; H_\mathrm{eff}(s) \Big\} =:
	U_\mathrm{eff}(t,t_0)
	$$}

\subsection{Validation of the Effective Hamiltonian}
\label{sec:validation}
Our main goal is to find a relatively simple $H_\mathrm{eff}$ that accurately describes the dynamics in a suitable “relevant” subspace. The dynamics taken outside this subspace is not important, so $H_\mathrm{eff}$ needs not to be accurate there. 
Since we are interested in the dynamics, it is natural to compare the exact population histories and coherences for the low-energy dynamics with those obtained with $H_\mathrm{eff}$.

More compact and effective quantifiers can be defined by adapting to our standard problem  metrics for operators in the Hilbert space, as the operatorial spectral norm or trace norm~\cite{zhao2021hamiltonian,an2021time,rajput2022hybridized,haah2021quantum,tran2019locality,berry2020time}. We anticipate that the effective Hamiltonian we will derive has a block diagonal structure, $H_\mathrm{eff}= P_0 \, H_\mathrm{eff} \, P_0 + (\mathds{1}-P_0)\, \, H_\mathrm{eff} \, (\mathds{1}-P_0)$, where the projection operator $P_0$ defines the relevant subspace. Since $[H_\mathrm{eff}, P_0]=0$ both the relevant subspace and its orthogonal complement are invariant under the effective dynamics. Then a suitable quantifier is defined as a fidelity

\begin{equation}
\label{eq:relevant-fidelity}
	F =  \min_{ \ket{ \psi_0 } } \left\{ |\bra{ \psi_0 } \mathcal{U}^{\dagger} \mathcal{U}_\mathrm{eff} \ket{ \psi_0 }|^2 \right\} 
\end{equation}

where $\ket{\psi_0} = P_0 \ket{\psi_0}$ is a vector belonging to the relevant subspace. This subspace fidelity can be smaller than one, either because $H_\mathrm{eff}$ is not accurate in describing the dynamics in the relevant subspace or because the exact dynamics determines leakage from the relevant subspace, with probability 
$L=\braket{\psi_0}{U^\dagger(t)\, [\mathds{1}-P_0] \, U(t)|\psi_0}$. Therefore, we could define another figure of merit characterizing procedures where leakage from the subspace has been excluded by post-selection 

\begin{equation}
\label{eq:postselection-fidelity}
    F^\prime =  \min_{ \ket{ \psi_0 } } \left\{ \left|{\bra{ \psi_0 } \mathcal{U}^{\dagger} P_0 \over \sqrt{1-L}} \, \mathcal{U}_\mathrm{eff} \ket{ \psi_0 }\right|^2 \right\} =\min_{ \ket{ \psi_0 } } \left\{ {|\bra{ \psi_0 } \, \mathcal{U}^{\dagger} \mathcal{U}_\mathrm{eff} \ket{ \psi_0 }|^2 \over 1-L}  \right\} 
\end{equation}

which is the subspace fidelity between the effective dynamics and the post-selected vector.

The impact of leakage will be quantified by approximating $F^\prime \approx F_m^\prime := F + L_m$, where $L_m$ is the probability of leakage evaluated for the initial state that enters the minimization determining $F$ in Equation (\ref{eq:relevant-fidelity}). This approximation is justified if both the infidelity and the leakage are small, $I := 1-F \ll 1$ and $L\ll 1 $, and arguing that while $F_m^\prime$ is not a lower bound as $F^\prime$ in Equation (\ref{eq:postselection-fidelity}), the worst-case error may be a significant overestimate for many initial states~\cite{zhao2021hamiltonian,an2021time}.
\section{Application to Adiabatic Elimination}
\label{section:ad_el_pr}
We now apply the procedure outlined to  a three-level system in
Lambda configuration modelled by the Hamiltonian

\begin{equation}
    \hat H = - \frac{\delta}{2} \ketbra{0}{0} + \frac{\delta}{2} \ketbra{1}{1} + \Delta \ketbra{2}{2} + \frac{1}{2} \sum_{k=0,1} \big[ \Omega_k^* \,\ketbra{k}{2} + \mbox{h.c.} \big].
    \label{eq:hamiltonian-lambda}
\end{equation}

  which describes a quantum network with on-site energies $(\pm \delta/2, \Delta/2)$ and tunnelling amplitudes $\Omega_k$ (see Figure \ref{fig:pops2}a). 
The same Hamiltonian provides the standard description in a rotating frame of a three-level atom driven by two near-resonant corotating semiclassical AC electromagnetic fields. In this case, $\Omega_k$ values represent the amplitudes of the fields while  the single-photon detunings between the atomic level splitting { $E_i-E_j$ quasi-resonant with the frequencies} $\omega_k$ of the fields,
$\delta_0 := E_2-E_0- \omega_0$ and $\delta_1 := E_2-E_1-\omega_1$,  enter the diagonal elements {as} $\delta := \delta_2 - \delta_1$ and $\Delta = (\delta_2+\delta_1)/2$. This model describes several three-level coherent phenomena used in quantum protocols from Raman oscillations~\cite{kr:01-vitanov-advatmolopt,ka:17-falci-fortphys-fqmt}, stimulated Raman adiabatic passage~\cite{kr:17-vitanovbergmann-rmp,ka:09-siebrafalci-prb,ka:19-falci-scirep-usc}, and hybrid schemes~\cite{ka:15-distefano-prb-cstirap,ka:19-falci-scirep-usc}.

The dynamics is governed by the Schr\"odinger equation $i \, \dot c_i(t) = \sum_{j=0}^{2} \bra{i} \hat H \ket{j} \, c_j(t)$ for 
$i,j = 0, \, 1,\,2 $. 
The system is prepared in the subspace spanned by the two lowest energy states, i.e., $\ket{\psi_0}= c_0 \ket{0} + c_1 \ket{1}$, which is defined as the "relevant" subspace. We want to understand  under which conditions the dynamics is confined to the relevant subspace, and to determine a Hamiltonian operator $\hat H_\mathrm{eff}$ whose projection onto this subspace effectively generates the confined dynamics.
\subsection{Adiabatic Elimination: Ambiguities and Limitations}
\label{sec:AE-ambiguities}
AE offers a simple and handy solution to this problem~\cite{kr:01-vitanov-advatmolopt}. The standard procedure~\mbox{\cite{ka:07-brion-jphysA-adiabelimination, ka:14-paulish-epjplus-adiabelim,kr:01-vitanov-advatmolopt}} relies on the observation that if $\Delta \gg \delta, | \Omega_k |$ for $k=0,1$, transitions from the lowest energy doublet and the state $\ket{2}$ are suppressed. 
Then, assuming that $\dot c_2(t)$ can be neglected in the Schr\"odinger equation, we find $ c_2 = - \tfrac{ \Omega_0 }{ 2 \Delta } \, c_0 - \tfrac{ \Omega_1 }{ 2 \Delta } \, c_1 $. Substituting in the equations for $\{c_0,c_1\}$ we obtain an effective two-level problem 
$i \partial_t \ket{\phi} = \hat H_\mathrm{eff} \ket{\phi}$ where 

$$
\hat{H}_\mathrm{eff} = -\Big( \frac{ \delta }{ 2 } - S_0  \Big) \ketbra{0}{0} + \Big( \frac{ \delta }{ 2 } + S_1  \Big) \ketbra{1}{1} + \Big( \frac{ \tilde{\Omega} }{ 2 } \ketbra{1}{0} + \text{h.c.} \Big)
$$

where $ S_k = - |\Omega_k|^2/(4 \, \Delta) $, $ k = 0, \, 1$ are energy shifts and $ \tilde{ \Omega } = - \Omega_0 \, \Omega^*_1/(2 \, \Delta ) $ is the normalized coupling. Since $\hat{H}_\mathrm{eff}$ is defined in the relevant subspace,  the state  $\ket{2} $ is not involved in the problem anymore. This procedure may be generalized to $ d > 3 $-level systems yielding an effective Hamiltonian in an $n < d$-dimensional relevant subspace. 

It has been pointed out in the literature 
~\cite{ka:07-brion-jphysA-adiabelimination,ka:14-paulish-epjplus-adiabelim} that standard AE suffers from a number of ambiguities and limitations, which we summarize here.
\begin{enumerate}
\item If we add to $H$ a term $\eta \mathds{1}$ that is an irrelevant uniform shift of all the energy levels, the procedure yields an $H_\mathrm{eff}$ that depends on $\eta$ in a non-trivial way. Thus, the procedure is affected by a gauge ambiguity. By comparing the exact numerical result with an analytic approximation based on the resolvent method a "best choice", $\eta =0$ has been proposed~\cite{ka:07-brion-jphysA-adiabelimination}.
\item AE completely disregards the state $\ket{2}$. However, although apparently confining the dynamics to the relevant subspace, the procedure yields that $ c_2(t) \neq 0 $ and depend on time. Thus, on the one hand, the approximation misses leakage to $\ket{2}$; on the other, it does not guarantee that the normalization of states of the relevant subspace is conserved. In Ref.~\cite{ka:14-paulish-epjplus-adiabelim}, the problem of normalization is overcome by writing separated differential equations in the relevant and in non-relevant subspaces.
\item The residual population in $\ket{2}$ as given by the approximate 
$|c_2(t)|^2$ may undergo very fast oscillations with angular frequency $\sim \Delta$. This is not consistent with the initial assumption that 
$\dot{c}_2 \approx 0 $. In Ref.~\cite{ka:14-paulish-epjplus-adiabelim}, the assumption is supported by arguing that it holds at the coarse-grained level, 
which averages out the dynamics at time-scales of $\sim \Delta^{-1}$ or faster.
\item Standard AE is not a reliable approximation for larger two-photon detunings or larger external pulses, and it is not clear how to systematically improve its validity. 
\end{enumerate}
We will show how the methodology outlined in Section~\ref{sec:effH-magnus} yields 
an effective formulation that overcomes the whole criticism above, leveraging only on coarse-graining of the dynamics.

\subsection{Magnus Expansion in the Regime of Large Detunings}
We now turn to coarse-graining via the ME. Keeping in mind the regime where 
$ \Delta \gg |\Omega_k| $ we first transform the Hamiltonian Equation (\ref{eq:hamiltonian-lambda}) to the interaction picture, 

$$
\tilde{H}(t) = U_0^\dagger \,(H - H_0)\, U_0= 
\frac{\Omega_0^*}{2} \, \ketbra{0}{2} \, e^{-i \left( \Delta + \delta/2 \right) t} + \, \frac{\Omega_1^*}{2} \, \ketbra{1}{2} \, e^{-i \left( \Delta - \delta/2 \right) t} + \text{h.c.}
$$

where $ U_0(t) := \mathrm{e}^{-i H_0 t} $ and $ H_0 $ is the diagonal part of $ H $. The first two terms of the ME in Equation (\ref{eq:magnus-terms}) are evaluated using the integrals reported in the Appendix~\ref{app:integrals}

$$
\begin{aligned}
\tilde{H}^{(1)}_\mathrm{eff}(t) &= \frac{ \Omega_0^*}{ 2 } \, e^{- i \left( \Delta + \delta/2 \right) t } \, \text{sinc} \left( \tfrac{ \left( \Delta + \delta/2 \right) \tau }{ 2 } \right) \, \ketbra{0}{2} + \\
&+ \, \frac{ \Omega_1^* }{ 2 } \, e^{- i \left( \Delta - \delta/2 \right) t } \, \text{sinc} \left( \tfrac{ \left( \Delta - \delta/2 \right) \tau }{ 2 } \right) \, \ketbra{1}{2} + \text{h.c.}
\\
\tilde{H}^{(2)}_\mathrm{eff}(t) &= {S}_0 \ketbra{0}{0} + {S}_1 \ketbra{1}{1} - (S_0+S_1) \ketbra{2}{2} + \frac{ \tilde{\Omega} }{2} \, \ketbra{1}{0} e^{ i \delta t } + \text{h.c.}
\end{aligned} 
$$

where the coefficients are given by

\begin{align}
{S}_{k} &= -\frac{ |\Omega_{k}|^2 }{4 \left( \Delta + (-1)^k \delta/2 \right)} \left[ 1 - \text{sinc} \left( \tfrac{ \Delta + (-1)^k \delta/2 }{ 2 } \tau \right) \right],
\\
\tilde{\Omega} &= - \frac{ \Omega_0 \Omega^*_1 }{ 2 } \, \frac{ \Delta }{ \left( \Delta^2 - \delta^4 / 4 \right) } \left[ \text{sinc} \tfrac{ \delta \tau }{ 2 } - \text{sinc} \tfrac{ \Delta \tau }{ 2 } \right],
\label{eq:eff-coupling0}
\end{align}
for $ k=0,1$. The first-order $\tilde{H}^{(1)}_\mathrm{eff}$ is an averaged version of $\tilde H$. The second-order $\tilde{H}^{(2)}_\mathrm{eff}$ contains shifts of the diagonal entries and an off-diagonal term coupling directly $\ket{0}$ and $\ket{1}$. At this order, that state $ \ket{2} $ is energy-shifted but not coupled to other states. 

\begin{figure}[H]
    \begin{subfigure}{.5\textwidth}
        \caption{}
        \centering
        \includegraphics[width=1\linewidth]{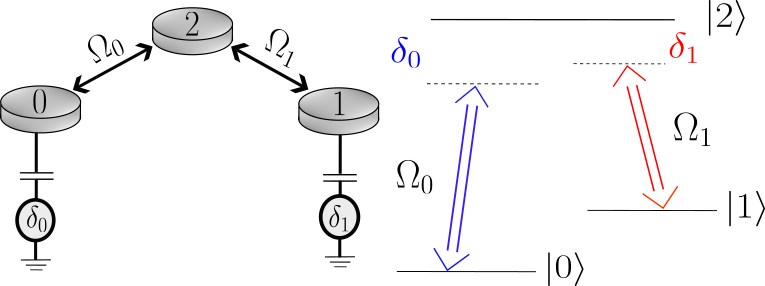}
    \end{subfigure}
    \begin{subfigure}{.5\textwidth}
        \caption{}
        \centering
        \includegraphics[width=1\linewidth]{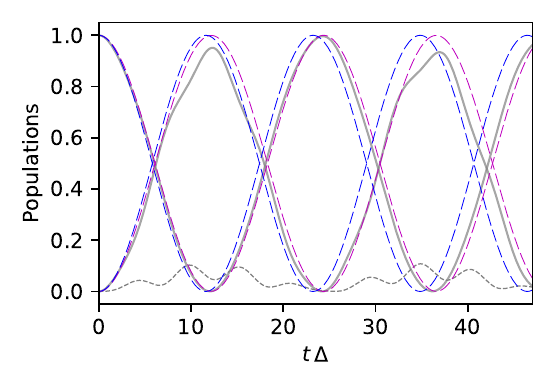}
    \end{subfigure}
    \begin{subfigure}{.5\textwidth}
        \caption{}
        \centering
         \includegraphics[width=1\linewidth]{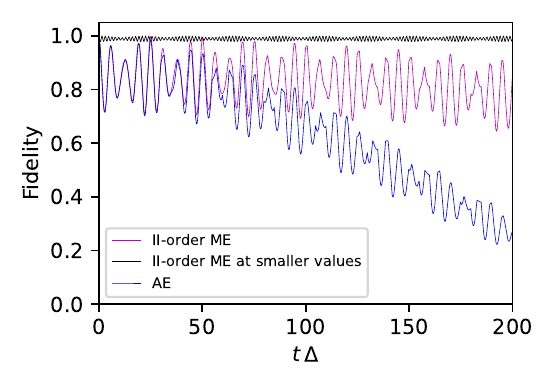}
    \end{subfigure}
    \begin{subfigure}{.5\textwidth}
        \caption{}
        \centering
        \includegraphics[width=1\linewidth]{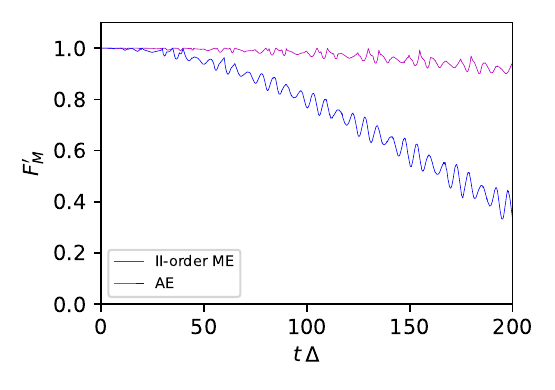}
    \end{subfigure}
    \protect \caption{
    {(\textbf{a})} The three-level model Equation (\ref{eq:hamiltonian-lambda}) describes, for instance, electronic levels confined in quantum dots $\Omega_k$ (on the left), representing tunnelling amplitudes and $\delta_i$ gate voltages to the ground; alternatively, it is the Hamiltonian in a rotating frame of a three-level atom (on the right) driven in $\Lambda$ configuration, $\Omega_k$ and $\delta_i$ being amplitudes and detunings of the external fields. {(\textbf{b}--\textbf{d})}      
    Comparison between exact dynamics  and effective dynamics obtained by ME (magenta) and standard AE (blue) for the parameters { values $\delta = 0.3 \, \Delta, \, \Omega_0 = 0.3 \, \Delta$ and $ \Omega_1 = 0.5 \Delta $}. {(\textbf{b})} 
    Population histories $|\braket{\psi_0 }{U_i(t)|\psi_0}|^2$ (curve starting from 1), $|\braket{\psi_1}{U_i(t)|\psi_0}|^2$ (starting from 0) and $|\braket{2}{U_i(t)|\psi_0}|^2$ (grey dashed). {The smooth grey lines refer to the exact evolution.} 
    {(\textbf{c})} The two-level fidelity $F$ \mbox{Equation (\ref{eq:relevant-fidelity})} at larger times. The black line refers to values of $(\delta, \Omega_k)$ four times smaller. {(\textbf{d})} The fidelity for protocols with post-selection $F^\prime_m=F+L_m$.}
    \label{fig:pops2}
\end{figure}

We now focus on the regime $ \Delta \gg \delta, \, |\Omega_k| $ where we can choose a coarse-graining time such that $2\pi/\Delta \ll \tau \ll 2\pi/\delta$. 
Then all the $\mathrm{sinc}(x)$ functions appearing in the terms of $H_\mathrm{eff}$ above are nearly vanishing except for $\mathrm{sinc}(\delta \tau /2) \approx 1$. As a result first-order $\tilde{H}^{(1)}_\mathrm{eff}(t)$ averages out while the coefficients of 
$\tilde{H}^{(2)}_\mathrm{eff}(t)$ become

\begin{align}
{S}_{k} = -\frac{ |\Omega_{k}|^2 }{4 \left( \Delta + (-1)^k \delta/2 \right)}
\quad ; \quad 
\tilde{\Omega} = - \frac{ \Omega_0 \Omega^*_1 }{ 2 } \, \frac{ \Delta }{ \left( \Delta^2 - \delta^4 / 4 \right) }
\end{align}

Transforming back to the laboratory frame, we finally obtain 

\begin{align}
    {H}_\mathrm{eff} = & - \left( \frac{ \delta }{ 2 } - S_0 \right) \ketbra{0}{0} + \left( \frac{ \delta }{ 2 } + S_1 \right) \ketbra{1}{1} \, + \, \left( \Delta \, - \, S_0 \, - \, S_1 \right) \ketbra{2}{2} \, + \nonumber
    \\
    &+ \, \frac{ \tilde{\Omega} }{ 2 } \, \ketbra{1}{0} \, + \, \text{h.c.}
    \label{Eq:effective_magnus}
\end{align}

Before discussing the results in detail, we make some general comments. As anticipated, we obtain a block-diagonal effective Hamiltonian. In the relevant subspace, it is similar to the result of standard AE. Concerning the criticism of the standard AE mentioned in Section~\ref{section:ad_el_pr}, we first observe that our result is not affected by the gauge ambiguity emerging for $H \to H + \eta \mathds{1}$, where the ``best choice'' rule $\eta =0$ of Ref.~\cite{ka:07-brion-jphysA-adiabelimination} is set naturally. Indeed, the uniform shift only changes trivially $U_0$ and does not enter $\tilde H$ where only level splittings appear. 
Secondly, the state $\ket{2}$ is not eliminated but the block diagonal structure consistently preserves normalization in each subspace. Thus, there is no need to assume that $\dot c_2=0$. Rather, Equation (\ref{Eq:effective_magnus}) may also describe  three-level dynamics where all coherences oscillate. Finally, the ME obviously allows for systematic improvement in the result of AE, which, moreover, can be extended significantly, as we will argue in the next sections.

While for $\delta = 0$ our $H_\mathrm{eff}$ is identical to the “best choice” result of the standard AE, differences emerge for $\delta \neq 0$. Anticipating the quantitative analysis of Section~\ref{sec:delta-neq-zero},  we here point out that our $H_\mathrm{eff}$ reproduces correctly the shifts $S_k$ as given by second-order perturbation theory (If $H$ describes a three-level atom under the action of corotating external fields $S_k$ are the perturbative Stark shifts), including the correct shift of the ``eliminated'' state $\ket{2}$. This feature is important for three-level of quantum operations, and an example will be discussed later. 

Coming back to coarse-graining, there is still another time scale to take into account. The solution may describe Raman oscillations of the populations in the relevant subspace with a period $\sim 2 \pi/| \tilde{\Omega} | $. Thus, $ \tau $ must be chosen that is small enough not to average out this dynamics while operating coarse-graining, which implies that $\Delta \gg \max(\delta,|\tilde{\Omega}|)$. In particular, for small $\delta$, we need  $\Delta \gg |\tilde{\Omega}|$, implying that $ | \Omega_0 \, \Omega_1 | /( 2 \, \Delta^2) \ll 1 $. Hence, in order for the approximation to work, we do not need both amplitudes $\Omega_k$ to be small separately, provided their product is small. Notice, finally, that if $\delta$ increases, say $\delta \gg |\tilde{\Omega}|$, but we still may choose the coarse-graining time as $\tau \gg \max \big(2 \pi/\Delta, 2\pi/\delta\big)$, then Equation (\ref{eq:eff-coupling0}) yields $\tilde{\Omega} \approx 0$. Differently from the standard AE, the three-level $H_\mathrm{eff}$ resulting by ME correctly decreases in this limit to the diagonal energy-shifted form obtained by perturbation theory.

\subsubsection{Comparison of the Results at the Second-Order Level} 
\label{sec:delta-neq-zero}
Besides overcoming the criticism raised to the standard AE, the ME approach  gives a good approximation already at the second-order level even when $| \Omega_0 \, \Omega_1 | /( 2 \, \Delta^2)$ is not very small. We here discuss  the case $ \delta \neq 0 $ where results of the second-order ME differ from those obtained by standard AE. We first look at the population histories considering the dynamics from an initial state of the relevant subspace $ \ket{\psi_0} = \cos \tfrac{ \theta^\prime }{ 2 } \ket{0} + \sin \tfrac{ \theta^\prime }{ 2 } \ket{1} $ with $ \theta^\prime = \theta + \tfrac{\pi}{2} $. For the curves in Figure~\ref{fig:pops2}b, the mixing angle $\theta^\prime$ is chosen such that $\ket{\psi_0}$ is an eigenstate of an observable { orthogonal to the effective Hamiltonian of the standard AE according to the Hilbert--Schmidt inner product defined in the Liouville vector space associated with the two-dimensional relevant subspace.} In other words, { if the effective AE Hamiltonian is proportional to a Pauli spin matrix forming an angle $\theta$ with $\sigma_z$, then we take a state with Bloch vector forming an angle $\theta^\prime = \theta + \pi/2$.} This choice is expected to nearly maximize the differences between the various cases.

Figure~\ref{fig:pops2}b shows that coarse-graining by ME yields population histories that  better approximate the exact dynamics in the relevant subspace. Notice that the discrepancy between ME and standard AE is very significant since it indicates a systematic error in the energies that accumulate over time. This clearly emerges from the fidelities $F$ defined in Equation (\ref{eq:relevant-fidelity}) 
and shown in Figure~\ref{fig:pops2}c where 
the minimization has been performed numerically. Indeed, $F$ becomes small for the standard AE, while the discrepancies between the ME result and the exact dynamics have a much smaller impact than what appears from the population histories. Actually, infidelity for the ME result is almost entirely due to leakage from the relevant subspace. These latter errors seem not to accumulate over time (magenta curve), as confirmed by the corresponding curve in Figure~\ref{fig:pops2}d, which shows that leakage errors can be corrected by post-selection. The residual error in the phase of the ME curves is remarkably small despite the exceedingly large value of $\sqrt{| \Omega_0 \, \Omega_1 | /( 2 \, \Delta^2)}\approx 0.27$ we used, and it is correctable by extending the analysis to the fourth order as we will show in the next section.
Errors due to leakage in Figure~\ref{fig:pops2}c almost disappear for values $\sqrt{| \Omega_0 \, \Omega_1 | /( 2 \, \Delta^2)} \ll 0.1$ (black curve in Figure~\ref{fig:pops2}c), and smaller values are even  routinely used for control of solid-state artificial atoms
. It is anyhow remarkable how accurate the ME coarse-grained second-order $H_\mathrm{eff}$ is in describing the protocol supplemented by post-selection for values of $ \delta $ and $ \Omega_k $ much beyond the perturbative regime
.
 
\subsubsection{Higher-Order Effective Hamiltonian} 
We now exploit the systematic improvement of the approximation. For the sake of simplicity, we will consider $ \delta=0 $. In this case, the exact eigenvalues can be calculated analytically, with the two splittings 
$\epsilon_{ij} := \epsilon_i-\epsilon_j$ being 
$\epsilon_{10} = \frac{\Delta}{2} \left( \sqrt{ 1 + 4 \, x } - 1 \right)$ 
and $\epsilon_{21} = \frac{\Delta}{2} \left( \sqrt{ 1 + 4 \, x } + 1 \right)$, 
where  $ x := \big(|\Omega_0|^2 + |\Omega_1|^2 \big)/( 4 \, \Delta^2)$. 
The second-order $ H_\mathrm{eff} $ in \mbox{Equation (\ref{Eq:effective_magnus})} reproduces the lowest-order expansion, $\epsilon_{10}/\Delta \approx x $ and $\epsilon_{21}/\Delta \approx 1+x$, and 
we now evaluate higher-order terms. Due to the algebra of the operators (see
Appendix~\ref{app:algebra}), the important property holds that terms of odd orders have the same structure as $\tilde{H}_\mathrm{eff}^{(1)}$, while at even orders they have the structure of $\tilde{H}_\mathrm{eff}^{(2)}$. In particular, the third-order term in the laboratory frame reduces to 

$$
{H}_\mathrm{eff}^{(3)} = \alpha (\tau)   \,x\, \left( \ketbra{2}{0} \frac{ \Omega_0 }{ 2 } + \ketbra{2}{1} \frac{\Omega_1 }{ 2 } \right)  + \mathrm{ h.c. }
$$

where $\alpha( \tau ) =
1 + { 1 \over 3 }\, \mathrm{sinc} \frac{ \Delta \, \tau }{ 2 }
\big[ 1 - 8 \, \cos (\Delta \, \tau / 2)  \big] $
up to an irrelevant factor of modulus one depending on the detailed coarse-graining procedure (see Appendix~\ref{app:integrals}). 

After coarse-graining over a time 
$ \tau \gg 2 \, \pi/\Delta \, $
we are left with $\alpha (\tau) \approx 1$. The resulting term triggers transitions between the relevant and the not relevant subspaces. While ${H}_\mathrm{eff}^{(3)}$ is $\sim x^{3/2}$, being off-diagonal contributes at order $x^3$ to the correction of the splittings, as  can be shown by ordinary non-degenerate perturbation theory. 

We turn to the fourth order of the ME, whose full expression is reported in \mbox{Appendix \ref{app:integrals}}. After coarse-graining, the contribution to the effective Hamiltonian in the laboratory frame reduces to

\begin{equation}
    H^{(4)}_\mathrm{eff} = S^{(4)}_0 \ketbra{ 0 }{ 0 } + S^{(4)}_1 \ketbra{ 1 }{ 1 } + ( \Delta - S^{(4)}_0 - S^{(4)}_1 ) \ketbra{ 2 }{ 2 } + \frac{ 1 }{ 2 } \,\big[ \tilde{\Omega}^{(4)} \, \ketbra{ 1 }{ 0 } + \text{h.c.} \big ]
\end{equation}

where $ S^{(4)}_k = x \,| \Omega_k |^2/(4 \Delta) $ and $ \tilde{\Omega}^{(4)} = x\,  \Omega_0 \, \Omega_1^*/(2 \Delta)$. 
This term reproduces the expansion of the exact splitting up to order $x^2$, providing a more relevant correction to $H^{(2)}_\mathrm{eff}$  \mbox{Equation (\ref{Eq:effective_magnus})} than the third-order $H^{(3)}_\mathrm{eff}$. Therefore, we can neglect this latter term and approximate $ H_\mathrm{eff} \approx  H^{(2)}_\mathrm{eff}+ H^{(4)}_\mathrm{eff}$. 

The above effective Hamiltonian reproduces both the exact energy splittings to order $x^2$. Being block-diagonal, it admits no leakage. We may wonder if including $H^{(3)}_\mathrm{eff}$ may yield useful information on leakage, but the answer is negative. Indeed, we checked that its impact on population histories is small. In particular, the ME 
yields a coarse-grained version of the population in $\ket{2}$, which is much smaller than the exact one, since this latter oscillates on a time scale of $\sim 2 \pi/\Delta$.

\subsubsection{Validation of the Results at Fourth-Order} 
\begin{figure}[H]
    \begin{subfigure}{.5\textwidth}
        \caption{}
        \centering
        \includegraphics[width=1\linewidth]{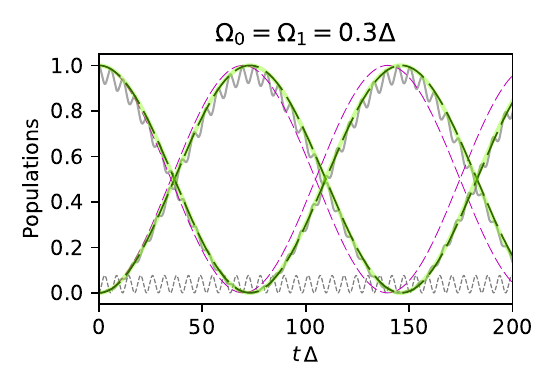}
    \end{subfigure}
    \begin{subfigure}{.5\textwidth}
        \caption{}
        \centering
        \includegraphics[width=1\linewidth]{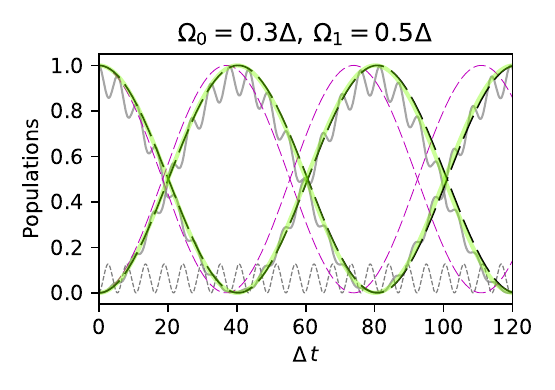}
    \end{subfigure}
    \caption{{(\textbf{a},\textbf{b})} Population histories $|\braket{\psi_0 }{U_i(t)|\psi_0}|^2$ (curve starting from 1), $|\braket{\psi_1}{U_i(t)|\psi_0}|^2$ (starting from 0), and $|\braket{2}{U_i(t)|\psi_0}|^2$ (grey dashed) for $\delta=0$. We compare for the exact dynamics (grey curves) and the higher-order approximate effective dynamics obtained by the second-order (magenta dashed) and fourth-order (black dashed)  ME. This latter coincides with the first-order Markov approximation of Ref.~\cite{ka:14-paulish-epjplus-adiabelim} (light green curve).}
    \label{fig:pops}
\end{figure}

\begin{figure}[H]
    \begin{subfigure}{.5\textwidth}
        \caption{}
        \centering
        \includegraphics[width=1\linewidth]{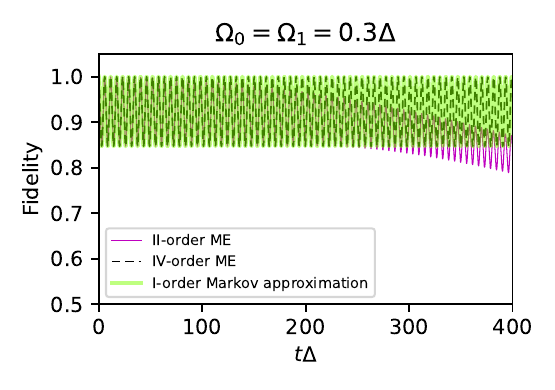}
    \end{subfigure}
    \begin{subfigure}{.5\textwidth}
        \caption{}
        \centering
        \includegraphics[width=1\linewidth]{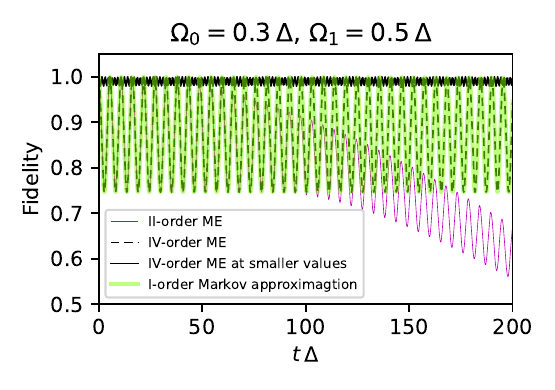}
    \end{subfigure}
        \begin{subfigure}{.5\textwidth}
        \caption{}
        \centering
        \includegraphics[width=1\linewidth]{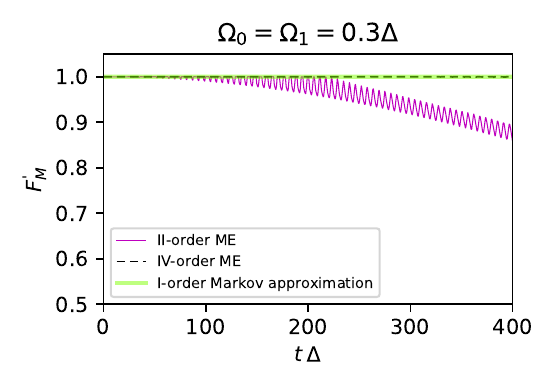}
      \end{subfigure}
      \begin{subfigure}{.5\textwidth}
        \caption{}
        \centering
        \includegraphics[width=1\linewidth]{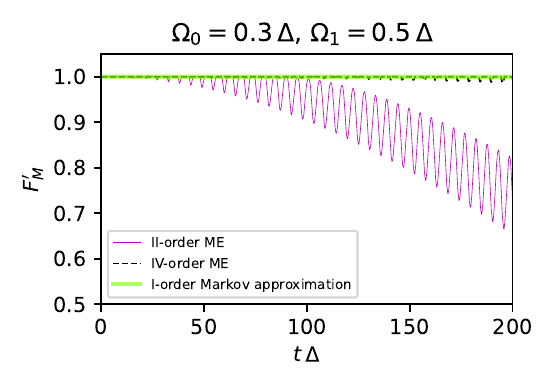}
    \end{subfigure}
    \caption{{(\textbf{a},\textbf{b})} Subspace fidelities $F$ for the parameters indicated. The curves refer to the second-order ME (magenta) and to the fourth-order ME (black dashed), which practically coincides with the Lippmann--Schwinger approach in Markov approximation of Ref.~\cite{ka:14-paulish-epjplus-adiabelim} (green line). 
    The second-order ME is affected by phase errors accumulating in time, whereas the error in the other two curves does not accumulate since it is due to leakage. In this latter case, $F$ is still large, considering the 
    values of the parameters are well beyond the perturbative regime. 
     The black solid line in panel (\textbf{b}) refers to values of $(\delta, \Omega_k )$ that are four times smaller, and the leakage error also  disappears. 
     {(\textbf{c},\textbf{d})} The fidelity for protocols with post-selection $F^\prime_m=F+L_m$. The fourth-order ME (black dashed) and the approximation of Ref.~\cite{ka:14-paulish-epjplus-adiabelim} (green line)
     are remarkably accurate also for values of the parameters well beyond the perturbative regime. }
    \label{fig:fids}
\end{figure}

We now validate our result using the same quantifiers as in  Section~\ref{sec:delta-neq-zero}. Population histories are shown in 
Figure~\ref{fig:pops} for two different sets of parameters. It is seen that the fourth-order ME (black dashed line) yields a coarse-grained version of the exact result (full grey line), which is accurate in reproducing the Raman oscillations between levels of the relevant subspace. On the contrary, the second-order ME (dashed magenta line) 
clearly shows a discrepancy in the oscillation frequency. This error 
appears clearly in the long-time fidelity $F$, shown in Figure~\ref{fig:fids}a,b, and the fidelity of post-selected protocols, 
shown in Figure~\ref{fig:fids}c,d. In the same figures, we also compare the fourth-order ME with the approximation scheme proposed in Ref.~\cite{ka:14-paulish-epjplus-adiabelim} (green curves), and the two approaches are seen to coincide. Therefore, the ME provides a systematic approximation scheme, overcoming the last point of the criticism to standard AE mentioned in  Section~\ref{sec:AE-ambiguities}. 

In particular for the symmetric choice $ |\Omega_0|= |\Omega_1| = 0.3 \, \Delta $ (Figure~\ref{fig:fids}a,c) the fourth-order fidelity oscillates between one and $ 0.8 $ (black dashed curve), whereas the second-order result (magenta dashed curve) decays to lower values because the error in frequencies accumulates in time. The same behaviour is obtained for the asymmetric configuration of the drive amplitudes shown in Figure~\ref{fig:fids}b,d. For these time scales, the error in the fourth-order $F$ in Figure~\ref{fig:fids}a,b does not accumulate in time, and Figure~\ref{fig:fids}c,d show that it is { entirely} due to leakage since it can be corrected by post-selection. Again, we notice that in \mbox{Figures \ref{fig:pops} and \ref{fig:fids}}, we used parameters with values far beyond the perturbative regime to magnify the errors. Still, errors are not so large, and in particular, they  are remarkably small for the post-selected dynamics. As in Figure~\ref{fig:pops2}c, errors due to leakage in Figure~\ref{fig:fids}b (black curve) almost disappear already for values $\sqrt{| \Omega_0 \, \Omega_1 | /( 2 \, \Delta^2)} \ll 0.1$.

Finally, we stress the agreement of the ME result with the results of the approach of Ref.~\cite{ka:14-paulish-epjplus-adiabelim}. This latter achieves a high-level accuracy using an iterated Lippmann--Schwinger equation in a special gauge defining the interaction picture and supplementing the problem by an ad hoc assumption of Markovianity of the dynamics. Our approach based on the ME shows that the correct result only leverages coarse-graining. 

\section{Discussion and Conclusions}
We introduced a technique based on the Magnus expansion that allows deriving a coarse-grained effective Hamiltonian and yielding a simplified model for the low-energy dynamics of the system. We applied the technique to the problem of the three-level lambda system clarifying ambiguities and inconsistencies of standard AE, which have been raised in the literature~\cite{ka:07-brion-jphysA-adiabelimination,ka:14-paulish-epjplus-adiabelim}.  

Results from ME accurately interpolate all the limiting cases obtained by standard approximations. For instance, the second-order ME yields a result reducing to the usual AE for $\delta \ll |\Omega_k|$ but reproducing the perturbative Stark shifts for $\delta \gg |\Omega_k|$. Moreover, the accuracy of the ME can be systematically increased, reaching a full agreement with other accurate approximation schemes such as the one developed in Ref.~\cite{ka:14-paulish-epjplus-adiabelim}. Since the ME-based approximation we propose only relies on coarse-graining, this latter is identified as the key ingredient underlying all the approximations. 

Coarse-grained  approximations yield effective Hamiltonians that accurately reproduce the dynamics in the relevant subspace for a wide range of parameters, much beyond the perturbation regime. 
In particular, the accuracy achieved for protocols supplemented by post-selection is is remarkably large. {It would be interesting to complement our approach with improved convergence methods as the Magnus--Taylor expansion method}~\cite{ka:20-zeuch-annalsphysics-exactrwa} {developed for the stroboscopic dynamics of two-state systems to understand whether an eternal}~\cite{ka:21-kazuya-pra-eternaladiabaticity,ka:22-kazuya-quantum-oneboundtorule} {effective Hamiltonian can be determined.}

Notice that the ME approach we described yields a block-diagonal effective Hamiltonian that does not cancel the not-relevant subspace but treats it consistently. In particular, for the Lambda system, we obtain the correct perturbative energy shift for the level $\ket{2}$. This result allows non-trivial applications to problems involving multiphoton processes in three-level dynamics~\cite{ka:17-falci-fortphys-fqmt,ka:16-vepsalainen-photonics-squtrit}, which are relevant for newly developed quantum 
hardware~\cite{PhysRevLett.120.150504,ka:21-haysdevoret-science-andreevspin,ka:20-pellegrino-commphys-gjj}. 

Finally, the approach can be extended to the design of effective control Hamiltonians
in time-dependent problems. The extension is simple if the parameters of the Hamiltonian are slowly varying on time scales of the order of the coarse-graining time $\tau$ as in Ref.~\cite{di2016coherent}. {Providing simpler and “slower” effective Hamiltonians, the ME can be used in numerical approaches}~\cite{BLANES2009151}, {especially with intensive algorithms such those based on optimal control theory}~\cite{KochEQT2022quantum} {or reinforcement learning}~\cite{ka:21-brown-njp-rlcoherentpop,ka:22-giannelli-pla-tutorialML}.

\vspace{6pt}
\authorcontributions{Conceptualization and methodology: N.M., L.G., E.P. and G.F.; software: N.M.; validation: N.M., L.G., E.P. and G.F.; formal analysis: N.M., L.G., E.P. and G.F.; writing---original draft preparation, N.M.; writing---review and editing: L.G., E.P. and G.F.; visualization: N.M., L.G.; supervision: E.P. and G.F.; project administration, G.F.; funding acquisition: E.P. and G.F. All authors have read and agreed to the published version of the manuscript.}

\funding{This work was supported by the QuantERA grant SiUCs (Grant No. 731473); by the University of Catania, Piano Incentivi Ricerca di Ateneo 2020-22, project Q-ICT; by the Partenariato Esteso “National Quantum
Science and Technology Institute (NQSTI)” - ambito di intervento “4. Scienze e Tecnologie Quantistiche”; and by ICSC---Centro Nazionale di Ricerca in High-Performance Computing, Big Data, and Quantum Computing, co-funded by European Union---NextGenerationEU.}

\institutionalreview{Not applicable.}

\dataavailability{Not applicable.} 

\acknowledgments{We acknowledge Kazuya Yuasa for the very interesting remarks, Dario	Zappalà and Giulia Termini for helpful discussions.
\\
This work was supported by the QuantERA grant SiUCs (Grant No. 731473), by the PNRR MUR project PE0000023-NQSTI, by ICSC – Centro Nazionale di Ricerca in High-Performance Computing, Big Data and Quantum Computing, by the University of Catania, Piano Incentivi Ricerca di Ateneo 2020-22, project Q-ICT. EP acknowledges the COST Action CA 21144 superqumap.}

\conflictsofinterest{The authors declare no conflicts of interest. The funders had no role in the design of the study; in the collection, analyses, or interpretation of data; in the writing of the manuscript; or in the decision to publish the results.} 

\abbreviations{Abbreviations}{
The following abbreviations are used in this manuscript:\\

\noindent 
\begin{tabular}{@{}ll}
ME & Magnus expansion\\
AE & Adiabatic elimination
\end{tabular}
}

\appendixtitles{yes} 
\appendixstart
\appendix
\section[\appendixname~\thesection]{More on the ME}
\subsection{Third and Fourth-Order Terms}
\label{app:ME_high_order}
We report here the higher-order terms we use in this work; see \cite{BLANES2009151} for more details. The third-order term is 

$$
F_3 =  \frac{ (-i)^2 }{ 6 }  \int_{t_0}^{t} \int_{t_0}^{s_1} \int_{t_0}^{s_2} \,
ds_1 \, ds_2 \, ds_3 \, \left\{ \comm{ H(s_1) }{ \comm{ H(s_2) }{ H(s_3) } } +
  \comm{ \comm{ H(s_1) }{ H(s_2) } }{ H(s_3) } \right\},
$$

while at the fourth order we have 

$$
\begin{aligned}
    F_4 = \frac{ (-i)^3 }{ 12 }  &\int_{t_0}^{t} \int_{t_0}^{s_1} \int_{t_0}^{s_2} \int_{t_0}^{s_3} \, ds_1 \, ds_2 \, ds_3 \, ds_4 \Bigl\{ \comm{ \comm{ \comm{ H(s_1) }{ H(s_2) } }{ H(s_3) } }{ H(s_4) } + \\
    &+ \comm{ H(s_1) }{ \comm{ \comm{ H(s_2) }{ H(s_3) } }{ H(s_4) } } + \comm{ H(s_1) }{ \comm{ H(s_2) }{ \comm{ H(s_3) }{ H(s_4) } } } + \\ 
    &+ \comm{ H(s_2) }{ \comm{ H(s_3) }{ \comm{ H(s_4) }{ H(s_1) } } } \Bigr\}.
\end{aligned}
$$

\subsection[\appendixname~\thesection]{Convergence}
\label{app:convergence}
The convergence of the Magnus expansion depends on the necessary condition

$$\int_{0}^\tau \lVert \tilde H \rVert ds < \pi$$

where the spectral norm is intended~\cite{BLANES2009151, casas2007sufficient}. This limit is a more {precise} version of $ \tau \lVert \tilde H \rVert \ll 1 $~\cite{brinkmann2016introduction} and it may be rather conservative. Indeed, the radius of convergence of the series may be larger~\cite{BLANES2009151}. For instance, for the Hamiltonian

$$ \tilde H \, (t) = \frac{ \Omega }{ 2 } \, \mathrm{e}^{-i \, \Delta \, t} \sum_{j<2} \ketbra{j}{2} + \mathrm{h.c}, $$

the above sufficient convergence condition becomes $ \tau < 2 \pi / \Omega $, and it gives a rather conservative bound on the coarse-graining time $\tau$.
{Actually, for practical
purposes, one might consider large values of $\tau$ yielding an asymptotic ME, which  can be proved to work up to a time increasing with the order}~\cite{ka:22-kazuya-quantum-oneboundtorule}.

\subsection[\appendixname~\thesection]{Structure of the series for the Lambda Hamiltonian}
The Lambda Hamiltonian we study has the structure

$$ \tilde H (s) = \sum_{i=0,1} V_i (s) \ketbra{i}{2} + \mathrm{h.c.} $$

with $V_i (s) = \Omega_i \, \mathrm{e}^{-i \, \epsilon_{2i} \, s} /2 $. The second-order term of ME contains the commutator 

\begin{equation}
    C_\mathrm{II} (s,s') = \comm{H (s)}{H (s')} = \sum_{i,j = 0,1} \, f_{i,j} (s,s') \, \left( \ketbra{2}{2} \, \delta_{i,j} - \ketbra{j}{i} \right)
    \label{eq:ev_ord}
\end{equation}

where $ f_{i,j}(s,s') = V^*_i ( s ) \, V_j ( s' ) - V_j ( s ) \, V^*_i ( s' ) $. These terms come from $ \comm{ \ketbra{ 2 }{ i } }{ \ketbra{ j }{ 2 } } $. This structure yields shifts on the diagonal elements of the Hamiltonian and connects the states of the relevant subspace (see Equation~\eqref{Eq:effective_magnus}).

Continuing with the third order, we encounter the nested commutator 

\begin{equation}
   C_\mathrm{III} (s,s',s'') = \comm{ \comm{ H(s) }{ H(s') } }{ H(s'') } = \sum_{i=0,1} g_i(s,s',s'') \ketbra{ i }{ 2 } + \mathrm{h.c.}
   \label{eq:odd_ord}
\end{equation}

where $ g_i(s,s',s'') = - \sum_{j=0,1} \big[ f_{i,i} (s,s') \, V_j (s'') + f_{i,j} (s,s') \, V_i (s'') \big] $. This quantity has {the} same structure of $\tilde H (s)$ and comes from $\comm{ \ketbra{ 2 }{ 2 } }{ \ketbra{ i }{ 2 } } = - \ketbra{ i }{ 2 } $ and $\comm{ \ketbra{ i }{ j } }{ \ketbra{ k }{ 2 } } = \delta_{j,k} \, \ketbra{i}{2} $. This argument can be iterated by showing that even (odd) nested commutators have the structure of Equation~\eqref{eq:ev_ord} (Equation~\eqref{eq:odd_ord}).
\label{app:algebra}
\section[\appendixname~\thesection]{Integrals}
In this work, the ME at order $n$ requires the evaluation of  time-ordered integrals of the form 

$$
        \int_{t-\tau/2}^{t+\tau/2} dt_n \ldots \int_{t-\tau/2}^{s_2} dt_1 \, \mathrm{e}^{i \, \sum_{j=1}^n \, \omega_j \, t_j} 
        = \mathrm{e}^{i \sum_{j=1}^n \, \omega_j \,t} \int_{-\tau/2}^{\tau/2} ds_n \ldots \int_{-\tau/2}^{s_2} ds_1 \, \mathrm{e}^{i \, \sum_{j=1}^n \, \omega_j \, s_j} 
$$ 

If we shift the origin of the interval by $ \tau / 2 $, the integral becomes

\begin{equation}
    \mathrm{e}^{ i \sum_{j=1}^n \omega_j \, (t - \tau/2) } \int_{0}^{\tau} ds_n \ldots \int_{0}^{s_2} ds_1 \, \mathrm{e}^{i \, \sum_{j=1}^n \, \omega_j \, s_j}. 
    \label{eq:int-magnus}
\end{equation}

The integral can be calculated at all orders, and we find

\begin{align}
    &\int_{0}^{\tau} ds_n \ldots \int_{0}^{s_2} ds_1 \, \mathrm{e}^{i \, \sum_{j=1}^n \, \omega_j \, s_j} = (-i)^n \sum_{k=1}^n A_k^n \left( \mathrm{e}^{ i \Omega_{k-1}^n \, \tau } - 1 \right) \\
    & \Omega_{k-1}^n = \sum_{j=n-(k-1)}^n \omega_{j}, \qquad \Omega_0^n = \omega_n \nonumber \\
    & A_k^n = \tfrac{ A_{k-1}^{n-1} }{ \Omega_{k-1}^n }, \qquad A_0^n = -\sum_{k=1}^n A_n^k, \qquad A_1^1 = \tfrac{ 1 }{ \omega_1 } \nonumber
\end{align}

We notice that the factor $ \mathrm{e}^{-i \sum_{j=1}^n \omega_j \, \tau/2 } $ in the integral Equation (\ref{eq:int-magnus}) depends on the choice of time slicing for coarse-graining. Indeed, if we choose to integrate between $[ t, t+\tau ]$ instead of $[ t-\tau/2, t+\tau/2 ]$, this factor does not appear. 
We expect that this dependence is irrelevant in most cases. Indeed, in our case, this dependence appears only in the third-order term of the ME, which we must consistently neglect in the order of accuracy of our calculations. 

Using the integrals above, we can calculate the fourth-order term of the ME, which is given by

$$
 {\tilde H}_\mathrm{eff}^{(4)} = \beta (\tau) \left\{ \frac{ {\tilde \Omega}^{(4)} }{2} \ketbra{0}{1} + \mathrm{h.c.} + \sum_{k=0}^1 S^{(4)}_k \, \ketbra{k}{k} - ( \sum_{k=0}^1 S^{(4)}_k ) \ketbra{2}{2} \right\}
$$

$$
\beta (\tau) = \left( 1 - \frac{ 2 }{ 3 } \, \cos \frac{ \Delta \, \tau }{ 2 } \, \mathrm{sinc} \frac{ \Delta \, \tau }{ 2 } - \frac{ 1 }{ 3 } \, \cos \Delta \, \tau \, \, \mathrm{sinc} \Delta \, \tau \right);
$$

$$
{\tilde \Omega}^{(4)} = \frac{ \Omega_0 \, \Omega_1^* }{ 2 \Delta } x;
$$

$$
S^{(4)}_k = \frac{ | \Omega_k |^2 }{ 4 \Delta } x.
$$

where $ x = \frac{ |\Omega_0|^2 + |\Omega_1|^2 }{ 4 \, \Delta^2 } $ is the small parameter. After choosing $\tau > 2 \pi / \Delta$, $ \beta (\tau) \to 1 $.
\label{app:integrals}

\begin{adjustwidth}{-\extralength}{0cm}

\reftitle{References}



\PublishersNote{}
%



\end{adjustwidth}

\begin{thebibliography}{999}

\bibitem[Dowling and Milburn(2003)]{dowling2003quantum}
Dowling, J.P.; Milburn, G.J.
\newblock Quantum technology: The second quantum revolution.
\newblock {\em Philos. Trans. R. Soc. Lond.  Ser. A Math. Phys. Eng. Sci.} {\bf 2003}, {\em
  361},~1655--1674.

\bibitem[Deutsch(2020)]{deutsch2020harnessing}
Deutsch, I.H.
\newblock Harnessing the power of the second quantum revolution.
\newblock {\em PRX Quantum} {\bf 2020}, {\em 1},~020101.

\bibitem[Koch \em{et~al.}()Koch, Boscain, Calarco, Dirr, Filipp, Glaser,
  Kosloff, Montangero, Schulte-Herbrüggen, Sugny, and
  Wilhelm]{KochEQT2022quantum}
Koch, C.P.; Boscain, U.; Calarco, T.; Dirr, G.; Filipp, S.; Glaser, S.J.;
  Kosloff, R.; Montangero, S.; Schulte-Herbrüggen, T.; Sugny, D.;  et~al.
\newblock Quantum Optimal Control in Quantum Technologies. {{Strategic}} Report
  on Current Status, Visions and Goals for Research in {{Europe}}.
\newblock \emph{EPJ Quantum Technol.} \textbf{2022}, {\em 9},~19.
\newblock {\url{https://doi.org/10.1140/epjqt/s40507-022-00138-x}}.

\bibitem[Nielsen and Chuang(2002)]{nielsen2002quantum}
Nielsen, M.A.; Chuang, I.
\newblock \emph{Quantum Computation and Quantum Information}; \newblock Cambridge University Press: Cambridge, UK, 2010.


\bibitem[Gisin and Thew(2007)]{gisin2007quantum}
Gisin, N.; Thew, R.
\newblock Quantum communication.
\newblock {\em Nat. Photonics} {\bf 2007}, {\em 1},~165--171.

\bibitem[Benenti \em{et~al.}(2009)Benenti, D'Arrigo, and
  Falci]{ka:09-benentifalci-prl-memorych}
Benenti, G.; D'Arrigo, A.; Falci, G.
\newblock Enhancement of Transmission Rates in Quantum Memory Channels with
  Damping.
\newblock {\em Phys. Rev. Lett.} {\bf 2009}, {\em 103},~020502.
\newblock {\url{https://doi.org/10.1103/PhysRevLett.103.020502}}.

\bibitem[Degen \em{et~al.}(2017)Degen, Reinhard, and
  Cappellaro]{degen2017quantum}
Degen, C.L.; Reinhard, F.; Cappellaro, P.
\newblock Quantum sensing.
\newblock {\em Rev. Mod. Phys.} {\bf 2017}, {\em 89},~035002.

\bibitem[Bongs \em{et~al.}(2019)Bongs, Holynski, Vovrosh, Bouyer, Condon,
  Rasel, Schubert, Schleich, and Roura]{bongs2019taking}
Bongs, K.; Holynski, M.; Vovrosh, J.; Bouyer, P.; Condon, G.; Rasel, E.;
  Schubert, C.; Schleich, W.P.; Roura, A.
\newblock Taking atom interferometric quantum sensors from the laboratory to
  real-world applications.
\newblock {\em Nat. Rev. Phys.} {\bf 2019}, {\em 1},~731--739.

\bibitem[Theis \em{et~al.}(2016)Theis, Motzoi, and
  Wilhelm]{ka:16-theis-pra-simgatesfreqcrowded}
Theis, L.; Motzoi, F.; Wilhelm, F.
\newblock Simultaneous gates in frequency-crowded multilevel systems using
  fast, robust, analytic control shapes.
\newblock {\em Phys. Rev. A} {\bf 2016}, {\em 93},~012324.

\bibitem[Li \em{et~al.}(2022)Li, Calarco, and
  Motzoi]{ka:22-li-prx-nonperturbativeandiag}
Li, B.; Calarco, T.; Motzoi, F.
\newblock Nonperturbative Analytical Diagonalization of Hamiltonians with
  Application to Circuit QED.
\newblock {\em PRX Quantum} {\bf 2022}, {\em 3},~030313.

\bibitem[Goerz \em{et~al.}()Goerz, Reich, and Koch]{GoerzNJP2014optimal}
Goerz, M.H.; Reich, D.M.; Koch, C.P.
\newblock Optimal Control Theory for a Unitary Operation under Dissipative
  Evolution.
\newblock \emph{New J. Phys.}
\textbf{2014}, {\em 16},~055012.
\newblock {\url{https://doi.org/10.1088/1367-2630/16/5/055012}}.

\bibitem[Blanes \em{et~al.}(2009)Blanes, Casas, Oteo, and Ros]{BLANES2009151}
Blanes, S.; Casas, F.; Oteo, J.; Ros, J.
\newblock The Magnus expansion and some of its applications.
\newblock {\em Phys. Rep.} {\bf 2009}, {\em 470},~151--238.

\bibitem[Brion \em{et~al.}(2007)Brion, Pedersen, and
  M{\o}lmer]{ka:07-brion-jphysA-adiabelimination}
Brion, E.; Pedersen, L.H.; M{\o}lmer, K.
\newblock Adiabatic elimination in a lambda system.
\newblock {\em J. Phys. Math. Theor.} {\bf 2007},
  {\em 40},~1033--1043.

\bibitem[Paulisch \em{et~al.}(2014)Paulisch, Rui, Ng, and
  Englert]{ka:14-paulish-epjplus-adiabelim}
Paulisch, V.; Rui, H.; Ng, H.K.; Englert, B.G.
\newblock Beyond adiabatic elimination: A hierarchy of approximations for
  multi-photon processes.
\newblock {\em  Eur. Phys. J. Plus} {\bf 2014}, {\em 129},~12.

\bibitem[Bravyi \em{et~al.}(2011)Bravyi, DiVincenzo, and
  Loss]{bravyi2011schrieffer}
Bravyi, S.; DiVincenzo, D.P.; Loss, D.
\newblock Schrieffer--Wolff transformation for quantum many-body systems.
\newblock {\em Ann. Phys.} {\bf 2011}, {\em 326},~2793--2826.

\bibitem[Bukov \em{et~al.}(2016)Bukov, Kolodrubetz, and
  Polkovnikov]{bukov2016schrieffer}
Bukov, M.; Kolodrubetz, M.; Polkovnikov, A.
\newblock Schrieffer-Wolff transformation for periodically driven systems:
  Strongly correlated systems with artificial gauge fields.
\newblock {\em Phys. Rev. Lett.} {\bf 2016}, {\em 116},~125301.

\bibitem[Blanes \em{et~al.}(2010)Blanes, Casas, Oteo, and
  Ros]{blanes2010pedagogical}
Blanes, S.; Casas, F.; Oteo, J.; Ros, J.
\newblock A pedagogical approach to the Magnus expansion.
\newblock {\em Eur. J. Phys.} {\bf 2010}, {\em 31},~907.

\bibitem[Petiziol \em{et~al.}()Petiziol, Arimondo, Giannelli, Mintert, and
  Wimberger]{PetiziolSR2020optimized}
Petiziol, F.; Arimondo, E.; Giannelli, L.; Mintert, F.; Wimberger, S.
\newblock Optimized Three-Level Quantum Transfers Based on Frequency-Modulated
  Optical Excitations. \emph{Sci Rep},
\newblock \textbf{2020}, {\em 10},~2185.
\newblock {\url{https://doi.org/10.1038/s41598-020-59046-8}}.

\bibitem[Vandersypen and Chuang(2005)]{vandersypen2005nmr}
Vandersypen, L.M.; Chuang, I.L.
\newblock NMR techniques for quantum control and computation.
\newblock {\em Rev. Mod. Phys.} {\bf 2005}, {\em 76},~1037.

\bibitem[Zhao \em{et~al.}(2021)Zhao, Zhou, Shaw, Li, and
  Childs]{zhao2021hamiltonian}
Zhao, Q.; Zhou, Y.; Shaw, A.F.; Li, T.; Childs, A.M.
\newblock Hamiltonian simulation with random inputs.
\newblock {\em arXiv} {\bf 2021}, arXiv:2111.04773.

\bibitem[An \em{et~al.}(2021)An, Fang, and Lin]{an2021time}
An, D.; Fang, D.; Lin, L.
\newblock Time-dependent unbounded Hamiltonian simulation with vector norm
  scaling.
\newblock {\em Quantum} {\bf 2021}, {\em 5},~459.

\bibitem[Rajput \em{et~al.}(2022)Rajput, Roggero, and
  Wiebe]{rajput2022hybridized}
Rajput, A.; Roggero, A.; Wiebe, N.
\newblock Hybridized methods for quantum simulation in the interaction picture.
\newblock {\em Quantum} {\bf 2022}, {\em 6},~780.

\bibitem[Haah \em{et~al.}(2021)Haah, Hastings, Kothari, and
  Low]{haah2021quantum}
Haah, J.; Hastings, M.B.; Kothari, R.; Low, G.H.
\newblock Quantum algorithm for simulating real time evolution of lattice
  Hamiltonians.
\newblock {\em SIAM J. Comput.} {\bf 2021}, pp. FOCS18--250.

\bibitem[Tran \em{et~al.}(2019)Tran, Guo, Su, Garrison, Eldredge, Foss-Feig,
  Childs, and Gorshkov]{tran2019locality}
Tran, M.C.; Guo, A.Y.; Su, Y.; Garrison, J.R.; Eldredge, Z.; Foss-Feig, M.;
  Childs, A.M.; Gorshkov, A.V.
\newblock Locality and digital quantum simulation of power-law interactions.
\newblock {\em Phys. Rev. X} {\bf 2019}, {\em 9},~031006.

\bibitem[Berry \em{et~al.}(2020)Berry, Childs, Su, Wang, and
  Wiebe]{berry2020time}
Berry, D.W.; Childs, A.M.; Su, Y.; Wang, X.; Wiebe, N.
\newblock Time-dependent Hamiltonian simulation with L1-norm scaling.
\newblock {\em Quantum} {\bf 2020}, {\em 4},~254.

\bibitem[Vitanov \em{et~al.}(2001)Vitanov, Fleischhauer, Shore, and
  Bergmann]{kr:01-vitanov-advatmolopt}
Vitanov, N.; Fleischhauer, M.; Shore, B.; Bergmann, K.
\newblock Coherent manipulation of atoms and molecules by sequential laser
  pulses.
\newblock {\em Adv.  At. Mol.  Opt. Phys.} {\bf 2001}, {\em 46},~55--190.
\newblock {\url{https://doi.org/10.1016/S1049-250X(01)80063-X}}.

\bibitem[Falci \em{et~al.}(2017)Falci, Di~Stefano, Ridolfo, D'Arrigo, Paraoanu,
  and Paladino]{ka:17-falci-fortphys-fqmt}
Falci, G.; Di~Stefano, P.; Ridolfo, A.; D'Arrigo, A.; Paraoanu, G.; Paladino,
  E.
\newblock Advances in quantum control of three-level superconducting circuit
  architectures.
\newblock {\em Fort. Phys.} {\bf 2017}, {\em 65},~1600077.
\newblock {\url{https://doi.org/10.1002/prop.201600077}}.

\bibitem[Vitanov \em{et~al.}(2017)Vitanov, Rangelov, Shore, and
  Bergmann]{kr:17-vitanovbergmann-rmp}
Vitanov, N.V.; Rangelov, A.A.; Shore, B.W.; Bergmann, K.
\newblock Stimulated Raman adiabatic passage in physics, chemistry, and beyond.
\newblock {\em Rev. Mod. Phys.} {\bf 2017}, {\em 89},~015006.
\newblock {\url{https://doi.org/10.1103/RevModPhys.89.015006}}.

\bibitem[Siewert \em{et~al.}(2009)Siewert, Brandes, and
  Falci]{ka:09-siebrafalci-prb}
Siewert, J.; Brandes, T.; Falci, G.
\newblock Advanced control with a Cooper-pair box: Stimulated Raman adiabatic
  passage and Fock-state generation in a nanomechanical resonator.
\newblock {\em Phys. Rev. B} {\bf 2009}, {\em 79},~024504.
\newblock 27 citazioni--da citare per h index totale,
  {\url{https://doi.org/10.1103/PhysRevB.79.024504}}.

\bibitem[Falci \em{et~al.}(2019)Falci, Ridolfo, Di~Stefano, and
  Paladino]{ka:19-falci-scirep-usc}
Falci, G.; Ridolfo, A.; Di~Stefano, P.; Paladino, E.
\newblock Ultrastrong coupling probed by Coherent Population Transfer.
\newblock {\em Sci. Rep.} {\bf 2019}, {\em 9},~9249.
\newblock {\url{https://doi.org/10.1038/s41598-019-45187-y}}.

\bibitem[Di~Stefano \em{et~al.}(2015)Di~Stefano, Paladino, D'Arrigo, and
  Falci]{ka:15-distefano-prb-cstirap}
Di~Stefano, P.G.; Paladino, E.; D'Arrigo, A.; Falci, G.
\newblock Population transfer in a Lambda system induced by detunings.
\newblock {\em Phys. Rev. B} {\bf 2015}, {\em 91},~224506.
\newblock {\url{https://doi.org/10.1103/PhysRevB.91.224506}}.

\bibitem[Zeuch \em{et~al.}(2020)Zeuch, Hassler, Slim, and
  DiVincenzo]{ka:20-zeuch-annalsphysics-exactrwa}
Zeuch, D.; Hassler, F.; Slim, J.J.; DiVincenzo, D.P.
\newblock Exact rotating wave approximation.
\newblock {\em Ann. Phys.} {\bf 2020}, {\em 423},~168327.

\bibitem[Burgarth \em{et~al.}(2021)Burgarth, Facchi, Nakazato, Pascazio, and
  Yuasa]{ka:21-kazuya-pra-eternaladiabaticity}
Burgarth, D.; Facchi, P.; Nakazato, H.; Pascazio, S.; Yuasa, K.
\newblock Eternal adiabaticity in quantum evolution.
\newblock {\em Phys. Rev. A} {\bf 2021}, {\em 103},~032214.

\bibitem[Burgarth \em{et~al.}(2022)Burgarth, Facchi, Gramegna, and
  Yuasa]{ka:22-kazuya-quantum-oneboundtorule}
Burgarth, D.; Facchi, P.; Gramegna, G.; Yuasa, K.
\newblock One bound to rule them all: From Adiabatic to Zeno.
\newblock {\em Quantum} {\bf 2022}, {\em 6},~737.

\bibitem[Veps\"al\"ainen \em{et~al.}(2016)Veps\"al\"ainen, Danilin, Paladino,
  Falci, and Paraoanu]{ka:16-vepsalainen-photonics-squtrit}
Veps\"al\"ainen, A.; Danilin, S.; Paladino, E.; Falci, G.; Paraoanu, G.S.
\newblock Quantum Control in Qutrit Systems Using Hybrid Rabi-STIRAP Pulses.
\newblock {\em Photonics} {\bf 2016}, {\em 3},~62.
\newblock {\url{https://doi.org/10.3390/photonics3040062}}.

\bibitem[Earnest \em{et~al.}(2018)Earnest, Chakram, Lu, Irons, Naik, Leung,
  Ocola, Czaplewski, Baker, Lawrence, Koch, and
  Schuster]{PhysRevLett.120.150504}
Earnest, N.; Chakram, S.; Lu, Y.; Irons, N.; Naik, R.K.; Leung, N.; Ocola, L.;
  Czaplewski, D.A.; Baker, B.; Lawrence, J.;  et~al.
\newblock Realization of a Lambda System with Metastable States of a
  Capacitively Shunted Fluxonium.
\newblock {\em Phys. Rev. Lett.} {\bf 2018}, {\em 120},~150504.
\newblock {\url{https://doi.org/10.1103/PhysRevLett.120.150504}}.

\bibitem[Hays \em{et~al.}(2021)Hays, Fatemi, Bouman, Cerrillo, Diamond,
  Serniak, Connolly, Krogstrup, Nyg{\aa}rd, Levy~Yeyati,
  et~al.]{ka:21-haysdevoret-science-andreevspin}
Hays, M.; Fatemi, V.; Bouman, D.; Cerrillo, J.; Diamond, S.; Serniak, K.;
  Connolly, T.; Krogstrup, P.; Nyg{\aa}rd, J.; Levy~Yeyati, A.;  et~al.
\newblock Coherent manipulation of an Andreev spin qubit.
\newblock {\em Science} {\bf 2021}, {\em 373},~430--433.

\bibitem[Pellegrino \em{et~al.}(2020)Pellegrino, Falci, and
  Paladino]{ka:20-pellegrino-commphys-gjj}
Pellegrino, F.; Falci, G.; Paladino, E.
\newblock 1/f critical current noise in short ballistic graphene Josephson
  junctions.
\newblock {\em Commun. Phys.} {\bf 2020}, {\em 3},~6.

\bibitem[Di~Stefano \em{et~al.}(2016)Di~Stefano, Paladino, Pope, and
  Falci]{di2016coherent}
Di~Stefano, P.; Paladino, E.; Pope, T.; Falci, G.
\newblock Coherent manipulation of noise-protected superconducting artificial
  atoms in the Lambda scheme.
\newblock {\em Phys. Rev. A} {\bf 2016}, {\em 93},~051801.

\bibitem[Brown \em{et~al.}(2021)Brown, Sgroi, Giannelli, Paraoanu, Paladino,
  Falci, Paternostro, and Ferraro]{ka:21-brown-njp-rlcoherentpop}
Brown, J.; Sgroi, P.; Giannelli, L.; Paraoanu, G.S.; Paladino, E.; Falci, G.;
  Paternostro, M.; Ferraro, A.
\newblock Reinforcement learning-enhanced protocols for coherent
  population-transfer in three-level quantum systems.
\newblock {\em N. J. Phys.} {\bf 2021}, {\em 23},~093035.

\bibitem[Giannelli \em{et~al.}(2022)Giannelli, Sgroi, Brown, Paraoanu,
  Paternostro, Paladino, and Falci]{ka:22-giannelli-pla-tutorialML}
Giannelli, L.; Sgroi, P.; Brown, J.; Paraoanu, G.S.; Paternostro, M.; Paladino,
  E.; Falci, G.
\newblock A tutorial on optimal control and reinforcement learning methods for
  quantum technologies.
\newblock {\em Phys. Lett. A} {\bf 2022}, \emph{2022}, 128054.
\clearpage
\bibitem[Casas(2007)]{casas2007sufficient}
Casas, F.
\newblock Sufficient conditions for the convergence of the Magnus expansion.
\newblock {\em J. Phys. Math. Theor.} {\bf 2007},
  {\em 40},~15001.

\bibitem[Brinkmann(2016)]{brinkmann2016introduction}
Brinkmann, A.
\newblock Introduction to average Hamiltonian theory. I. Basics.
\newblock {\em Concepts Magn. Reson. Part A} {\bf 2016}, {\em
  45},~e21414.

\end{thebibliography}
\end{document}